# Extraction of Text from Optic Nerve Optical Coherence Tomography Reports

Iyad Majid[1], Youchen (Victor) Zhang, MS[1], Robert Chang, MD[1], Sophia Y. Wang, MD, MS[1].
[1]Department of Ophthalmology, Byers Eye Institute, Stanford University, California, USA

## Abstract

**Purpose:** The purpose of this study was to develop and evaluate rule-based algorithms to enhance the extraction of text data, including retinal nerve fiber layer (RNFL) values and other ganglion cell count (GCC) data, from Zeiss Cirrus optical coherence tomography (OCT) scan reports.
**Methods:** DICOM files that contained encapsulated PDF reports with RNFL or Ganglion Cell in their document titles were identified from a clinical imaging repository at a single academic ophthalmic center. PDF reports were then converted into jpeg image files and processed using the open-source PaddleOCR Python package for optical character recognition. Rule-based algorithms were designed and iteratively optimized for improved performance in extracting RNFL and GCC data. Evaluation of the algorithms was conducted through manual review of a set of RNFL and GCC reports.
**Results:** The developed algorithms demonstrated high precision in extracting data from both RNFL and GCC scans. Precision was slightly better for the right eye in RNFL extraction (OD: 0.9803 vs. OS: 0.9046), and for the left eye in GCC extraction (OD: 0.9567 vs. OS: 0.9677). Some values presented more challenges in extraction, particularly clock hours 5 and 6 for RNFL thickness, and signal strength for GCC.
**Conclusions:** A customized optical character recognition algorithm can identify numeric results from optical coherence scan reports with high precision. Automated processing of PDF reports can greatly reduce the time to extract OCT results on a large scale.

## Introduction

Glaucoma, a complex and chronic eye disease, poses a significant global health challenge due to its insidious nature and potential for irreversible vision loss; the condition develops silently, with patients experiencing no noticeable symptoms until after significant visual field loss has occurred.[1,2] As a result, early detection and timely treatment are crucial in preventing irreversible vision loss in glaucoma patients.

Optical coherence tomography (OCT) scans have revolutionized the evaluation of ocular diseases including glaucoma.[3] OCT reports provide high-resolution cross-sectional images of the retina, allowing for precise measurement of retinal nerve fiber layer thickness (RNFLT), optic nerve head (ONH) parameters, and macular characteristics (including ganglion cell count). The integration of OCT scans into clinical practice has significantly improved the early detection of glaucoma and allows clinicians to better monitor disease progression with remarkable precision.[4] Consequently, OCT scans play a pivotal role in enhancing diagnostic capabilities and guiding treatment decisions for ophthalmologists. However, the wealth of information contained within OCT reports presents a challenge to extracting meaningful data for large-scale research studies. In addition, many researchers only have access to the clinical reports generated from OCT studies rather than direct access to all of the OCT scans' raw data due to the limitations of proprietary data formats specified by manufacturers.[5]

As OCT machines continue to improve and become more widespread, the volume of scans being generated has increased dramatically. According to Swanson and Fujimoto, about 30 million OCT scans are performed per year. The abundance of data makes manual management and analysis of data for research studies challenging, leading to a need for efficient methods in data extraction.[6] With the advent and rapid advancement of artificial intelligence (AI), text found in clinical data can now be easily extracted by improved optical character recognition (OCR) technology. These models allow for the extraction of data from OCT scans to better monitor the onset and progression of ocular diseases. The purpose of this study was to develop and validate algorithms to extract data from reports generated by Cirrus OCT scans of the retinal nerve fiber layer (RNFL) and the ganglion cell complex (GCC) data, so that results can be extracted on a large scale to facilitate future research in glaucoma.

# Methods

All DICOM files which contained encapsulated PDF reports where the document title contained either "RNFL" or "Ganglion Cell" were identified from the clinical imaging vendor neutral archive at a single academic center (Stanford Byers Eye Institute). The Pydicom Python package was used to read the DICOM headers to identify these files. These reports corresponded to optic nerve scans performed by Zeiss Cirrus OCT machines. A total of 22,506 RNFL reports and 22,022 GCC reports were identified. PDF reports were then extracted, converted, and converted into image files (JPG). The open source PaddleOCR (Optical Character Recognition) Python package was used as a starting point to detect text fields, recognize text characters from these fields, and then translate the text characters from the images into actual text strings. An advantage of PaddleOCR is that it has the ability to recognize text in "natural" contexts, that is to say, even when the text is embedded within a larger image. Our labeling of the RNFL clock hours from the RNFL reports is depicted in **Figure 1.** We refer to this scheme when discussing accuracy of extracted RNFL thickness values for each clock hour. An example of PaddleOCR output for RNFL and GCC reports is shown in **Figure 2**.

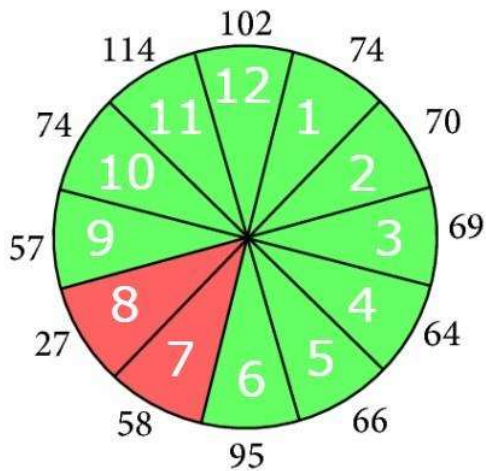

**Figure 1. Labeling for RNFL OCT Clock Hours.**
Legend: Labeling scheme of the RNFL clock hour segments from the report.

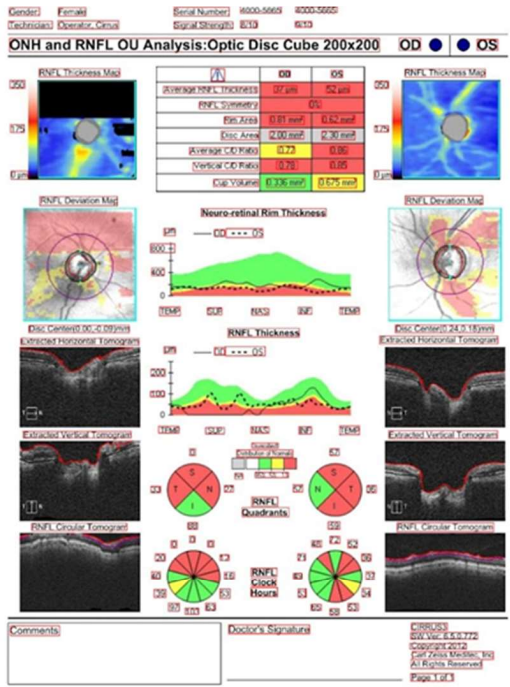

**Figure 2. Detection and highlighting of text fields from RNFL and GCC reports using PaddleOCR.**
Legend: Example RNFL and GCC reports are shown with red boxes around text that is highlighted by the PaddleOCR Python package. A selection of the PaddleOCR output is shown to the right of each report, indicating the detected text as well as the probability value for that detected text.

Due to the visual complexity of the imaging reports with some variation in length and placement of characters, OCR outputs had some inconsistencies in producing the same format and order of the results for each report when PaddleOCR was applied to the entire image. Therefore, part of the preprocessing algorithm crops relevant areas into separate images prior to passing these into the OCR algorithm. Then, rule-based algorithms were developed and applied for the RNFL and GCC OCR output to arrive at the final extracted data. We randomly sampled RNFL and GCC reports with a fixed random seed to iteratively tweak these two rule-based algorithms. For evaluation purposes, we randomly selected a final held-out evaluation set of 149 RNFL reports and 114 GCC reports for manual review of the performance of the final version of the RNFL and GCC rule-based algorithms. For each relevant text field on the report, we evaluated how many times the result was detected by the OCR algorithm, and of those detected outputs, how many were correct (positive predictive value, or precision). A public repository with code for this project is available at https://github.com/eyelovedata/oct-report-ocr.

# Results

Overall, the algorithm was highly precise in extracting results from RNFL (**Table 1**) and GCC (**Table 2**) reports. In general, the algorithm performed nearly equally on both eyes; on average, the algorithm was only slightly more precise when extracting RNFL data from the right eye (OD: 0.9803 vs. OS: 0.9046) and when extracting GCC data from the left eye (OD: 0.9567 vs. OS: 0.9677). Many fields for both RNFL and GCC were extracted with 100% precision. For RNFL, aside from RNFL thicknesses for individual clock hours, all of the main findings were extracted with greater than 0.95 precision. Results for RNFL thickness in individual clock hours varied, with clock hours 5 and 6 having the lowest precision of all the clock hours (in the 0.5 range), and clock hour the highest (0.9932). Extraction of GCC values was also generally highly precise, with the worst performing value extracted being the signal strength in the right eye (0.86). In general, the algorithm only occasionally failed to detect a result at all, mainly for individual clock hours on the RNFL report.

Several common patterns of OCR mistakes were evident upon manual assessment of the algorithm. The first was that the algorithm sometimes completely misread a data value, i.e the true value is 98 and the model reads it as 47. Another common mistake was the flipping of OD and OS values; for instance, if the cup-to-disk (C/D) ratio for OD and OS were 0.76 and 0.8 respectively, the algorithm extracts the values as 0.8 and 0.76. We also noticed our algorithm was sometimes one increment ahead of a sequence, i.e. it extracts the value for clock hour 2 and submits it as the value for clock hour 1. Finally, the data value itself was flipped during extraction, either horizontally (i.e. 86 is read as 68) or vertically (i.e. 66 is read as 99).

Table 1. OCR algorithm performance for RNFL OCT reports.

|  | OD | | OS | |
| --- | --- | --- | --- | --- |
|  | Precision | Total number of detected entries | Precision | Total number of detected entries |
| Signal Strength | 1 | 149 | 1 | 149 |
| Avg. RNFL Thickness | 1 | 149 | 1 | 149 |
| RNFL Symmetry | 1 | 149 | 0.9530 | 149 |
| Rim Area | 1 | 149 | 1 | 149 |
| Disc Area | 1 | 149 | 1 | 149 |
| Avg. C/D ratio | 0.9933 | 149 | 1 | 149 |
| Vertical C/D ratio | 0.9530 | 149 | 0.9799 | 149 |
| Cup Volume | 1 | 149 | 1 | 149 |
| Superior Quadrant | 0.9799 | 149 | 0.9645 | 142 |
| Temporal Quadrant | 0.9799 | 149 | 0.9799 | 149 |
| Nasal Quadrant | 0.9866 | 149 | 0.9866 | 149 |
| Inferior Quadrant | 0.9664 | 149 | 0.9728 | 145 |
| Clock Hour 1 | 0.7517 | 149 | 0.8993 | 149 |
| Clock Hour 2 | 0.8389 | 149 | 0.9762 | 119 |
| Clock Hour 3 | 0.8792 | 149 | 0.9762 | 119 |
| Clock Hour 4 | 0.9932 | 145 | 0.9933 | 149 |
| Clock Hour 5 | 0.5705 | 149 | 0.5588 | 140 |
| Clock Hour 6 | 0.5503 | 149 | 0.5436 | 149 |
| Clock Hour 7 | 0.9463 | 149 | 0.5570 | 149 |
| Clock Hour 8 | 0.9597 | 149 | 0.9799 | 149 |
| Clock Hour 9 | 0.9195 | 149 | 0.8792 | 149 |
| Clock Hour 10 | 0.8255 | 149 | 0.8322 | 149 |
| Clock Hour 11 | 0.7651 | 149 | 0.8792 | 149 |
| Clock Hour 12 | 0.9396 | 149 | 0.7987 | 149 |

Table 2. OCR algorithm performance for GCC OCT reports.

| | OD | | OS | |
|---|---|---|---|---|
| | Precision | Total number of detected entries | Precision | Total number of detected entries |
| Signal Strength | 0.8684 | 114 | 0.9912 | 114 |
| Superior Quadrant | 0.9912 | 114 | 0.9469 | 113 |
| Superior-nasal Quadrant | 0.9298 | 114 | 0.9469 | 113 |
| Inferior-nasal Quadrant | 0.9561 | 114 | 0.9646 | 113 |
| Inferior Quadrant | 0.9561 | 114 | 0.9813 | 107 |
| Inferior-temporal Quadrant | 0.9912 | 114 | 0.9642 | 112 |
| Superior-temporal Quadrant | 0.9825 | 114 | 0.9646 | 113 |
| Avg. GCL & IPL Thickness | 0.9561 | 114 | 0.9474 | 114 |
| Min. GCL & IPL Thickness | 0.9813 | 107 | 1 | 108 |

# Discussion

In this study, we developed and evaluated algorithms to extract clinical results from reports generated by Cirrus RNFL and GCC scans using optical character recognition. Using the open-source OCR package PaddleOCR as a base, we developed two algorithms to enhance performance in extracting text data from RNFL and GCC OCT scans. Evaluation of these algorithms on a sample of OCT reports showed that extraction was highly precise for GCC results and for most RNFL results, aside from the RNFL thicknesses of certain individual clock hours.

Because important results from clinical studies are often available to researchers only in PDF reports, building algorithms to extract this data simply and efficiently is key to enabling large-scale research efforts which can involve tens or hundreds of thousands of test results. Our study extends previous work to extract important results from a variety of other types of reports. One previous study developed an algorithm that extracts RNFL OCT data from Heidelberg Spectralis scans, a machine with a different type of report format compared to the Zeiss Cirrus scans which were the focus of the present efforts.[7] In addition to extracting numeric values, their system also uses deep learning to classify the findings on the images shown in the reports. Kim et al. developed several pre- and post-processing scripts in order to apply the Tesseract OCR engine to VF reports.[8] A similar study developed an open-source algorithm to extract data from Humphrey visual field reports.[9]

Human manual extraction is time-consuming, costly, labor-intensive, repetitive, and typically requires full concentration. It took our algorithm less than five minutes to process and extract data from 100 reports, but it may take hours over several days if done manually. Previous work on extraction of visual field data found that computers were 50 to 100 times faster at data extraction than humans.[9] Our algorithm may run effortlessly in the background of a computer while a human must dedicate time and energy to accurately extract the data, yet our algorithm's performance is comparable to that of human manual extraction.

Although our rule-based algorithm demonstrated high precision in extracting GCC OCT values, there were specific limitations in extracting the RNFL thicknesses from individual clock hours, particularly from clock hours 5 and 6 in both eyes. We found this was likely due to the irregular placement of the text, where the numbers are packed closely together but often offset from each other in different ways depending on the length of the numbers. Thus, we do not recommend using this approach for analyzing RNFL thicknesses for individual clock hours. However, in practice, these clock hour results are often more highly variable, and many clinicians rely more heavily on quadrant thicknesses which were extracted with high performance by our algorithm.[10]

In summary, we have developed a system to efficiently extract key parameters from Zeiss Cirrus reports for retinal nerve fiber layer (RNFL) thickness and ganglion cell count (GCC) OCT scans using optical character recognition. Our reported advancements have the potential to enhance the ability to collect large glaucoma clinical datasets efficiently and perform big data studies in glaucoma more quickly.